\documentclass[epj]{webofc}
\usepackage[varg]{txfonts}   % Web of Conferences font
\usepackage{subfigure}
\usepackage{times,mathptm}
\usepackage{amsmath,amsfonts}
\usepackage{mathptmx}
\usepackage{mathrsfs}
\usepackage{bbm}
\newcommand{\bea}{\begin{eqnarray}}
\newcommand{\eea}{\end{eqnarray}} 
\newcommand{\beq}{\begin{equation}}
\newcommand{\eeq}{\end{equation}}

\newcommand{\tth}{\widetilde{\theta}}

\newcommand{\tr}{\text{Tr}}

\newcommand{\bx}{\mathbf{x}}

\newcommand{\vx}{\bx}

\renewcommand{\b}{\beta}

\newcommand{\m}{\mu}

\newcommand{\s}{\sigma}

\newcommand{\oh}{\frac{1}{2}}

\newcommand{\dg}{\dagger}
\newcommand{\non}{\nonumber}

\newcommand{\rf}[1]{(\ref{#1})}
\newcommand{\ra}{\rightarrow}

\wocname{EPJ Web of Conferences}
\woctitle{CONF12}
\begin{document}
\selectlanguage{english}
\title{Confinement from Center Vortices}
%
% subtitle (optional, strongly discouraged)
%
\subtitle{A review of old and new results}

\author{Jeff Greensite\inst{1}\fnsep\thanks{Work supported by the US Department of Energy under Grant No.\
DE-SC0013682. \email{greensit@sfsu.edu}}}

\institute{Physics and Astronomy Dept., San Francisco State University, San Francisco CA 94132 USA}

\abstract{%
I briefly review the numerical evidence, some old and some quite recent, in favor of the center vortex
theory of confinement.
}
\maketitle
\section{Introduction}
\label{intro}
  
     In the absence of a proof, there is still no general consensus on the mechanism of quark confinement.
Is it monopoles, dyons, center vortices, the Gribov-Zwanziger scenario, or something else?  In my own opinion,
the weight of evidence favors center vortices (although connections to other ideas are not excluded), 
and I would like to review some of that evidence here.  Much of
what I'll present in this talk is very old, decades old in fact, and is found in various reviews \cite{Greensite:2011zz,Greensite:2003bk,Engelhardt:2004pf}.  Some of it, however, particularly the work
of Kamleh, Leinweber, and Trewartha \cite{Trewartha:2015nna,Trewartha:2015ida}, is quite new and, I believe, very persuasive.
  
    Let me begin with some motivation.  First consider Wilson loops in some color representation $r$ in a pure SU(N) gauge theory.  
Then the asympotic string tension $\s_r$ depends only on the N-ality of the representation, i.e.\ how the loop transforms under the
$Z_N$ center subgroup of the gauge group, and not on the representation itself.  There is an obvious reason for this: it is energetically favorable, for distant quark-antiquark sources in higher-dimensional representations,  for gluons to pop out of the vacuum and bind to the color sources,
effectively screening the color charge to the lowest dimensional representation of the same N-ality.  But color screening by gluons is a ``particle'' explanation.  What if we ask for a ``field'' explanation?  After all, the expectation value of a Wilson loop is obtained by summing over field configurations in a path integral.  What sort of field configurations have the property that they lead to an asymptotic string tension which depends only on N-ality?  As far as I know, the only configurations of this kind are center vortices.

    Of course the word ``confinement''  is itself a little controversial; what does it really mean?  Often people take it to mean that asymptotic particle states are colorless, and that is fine, providing one understands that gauge-Higgs theories (with the Higgs in the fundamental representation) also fall into this category.  In that case there are no flux tubes, no linear Regge trajectories, only Yukawa forces, and yet the asymptotic states are color neutral \cite{Fradkin:1978dv,Frohlich:1981yi}.  I prefer to reserve the word ``confinement'' to refer to an asymptotic property of the vacuum, when there are vacuum fluctuations that are so large that Wilson loops fall off with an asymptotic area law.  This terminology has a price, i.e.\ QCD becomes a ``confinement-like'' theory, but I think it is the cleanest way to distinguish the confinement phase of a non-abelian gauge theory from other phases that might exist.  If we define confinement in that way, then confinement is distinguished from other phases by a symmetry: {\it confinement is the phase of unbroken global center symmetry}.  Thus, in gauge-Higgs theories with the Higgs in an adjoint representation, the Higgs phase is a phase of spontaneously broken center symmetry.  In pure gauge theory at finite temperature, the
deconfined phase is also a phase of spontaneously broken center symmetry.  In QCD with light fermions in the fundamental representation, the global center symmetry is broken explicitly. In $G_2$ gauge theory, the center is only the identity element.  But whether the symmetry is broken spontaneously, or explicitly, or was trivial from the start, the absence of a non-trivial center symmetry always implies that the vacuum does not support confinement as I have defined it.  Since a global symmetry is involved, there are of course order parameters which distinguish the confined and unconfined phases.  These are the 't Hooft loop operator, Polyakov lines, and the center vortex free energy.  All of them refer to the global center symmetry of the action.

\begin{figure}[h!]
\centering
\subfigure[]  % caption for subfigure a
{   
 \label{global}
 \includegraphics[scale=0.4]{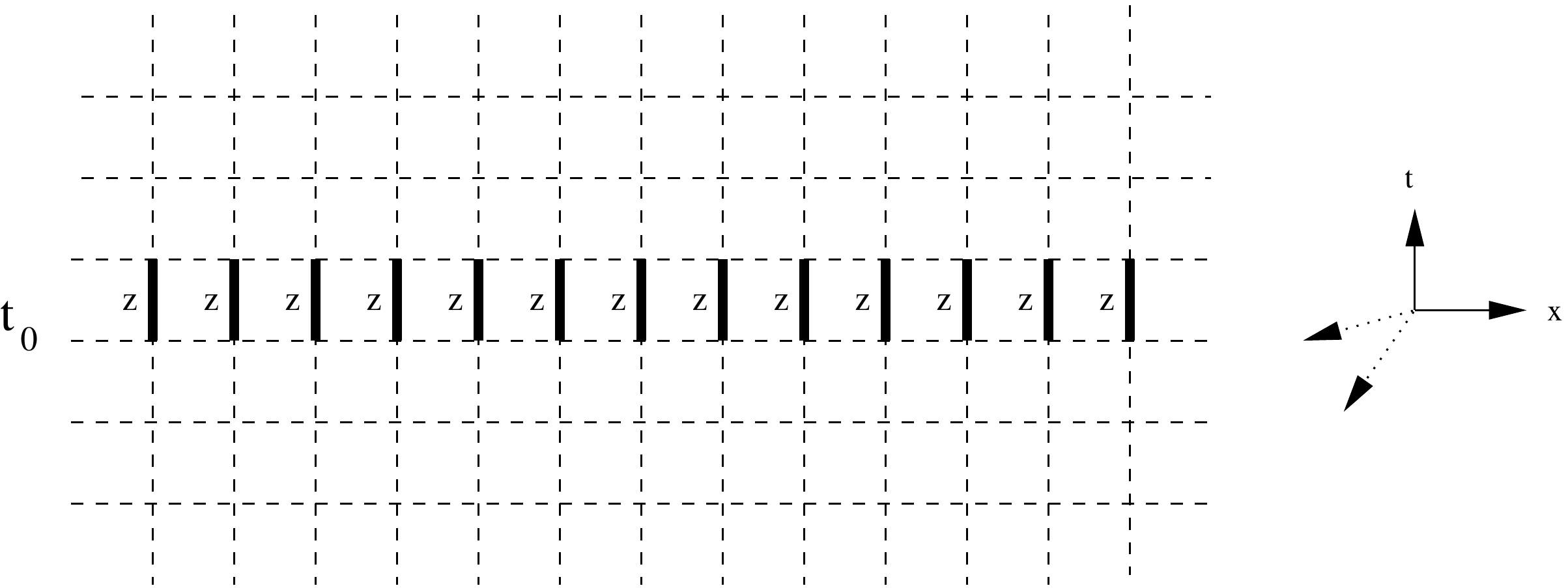}
}
\subfigure[]  % caption for subfigure a
{   
 \label{cv3}
 \includegraphics[scale=0.4]{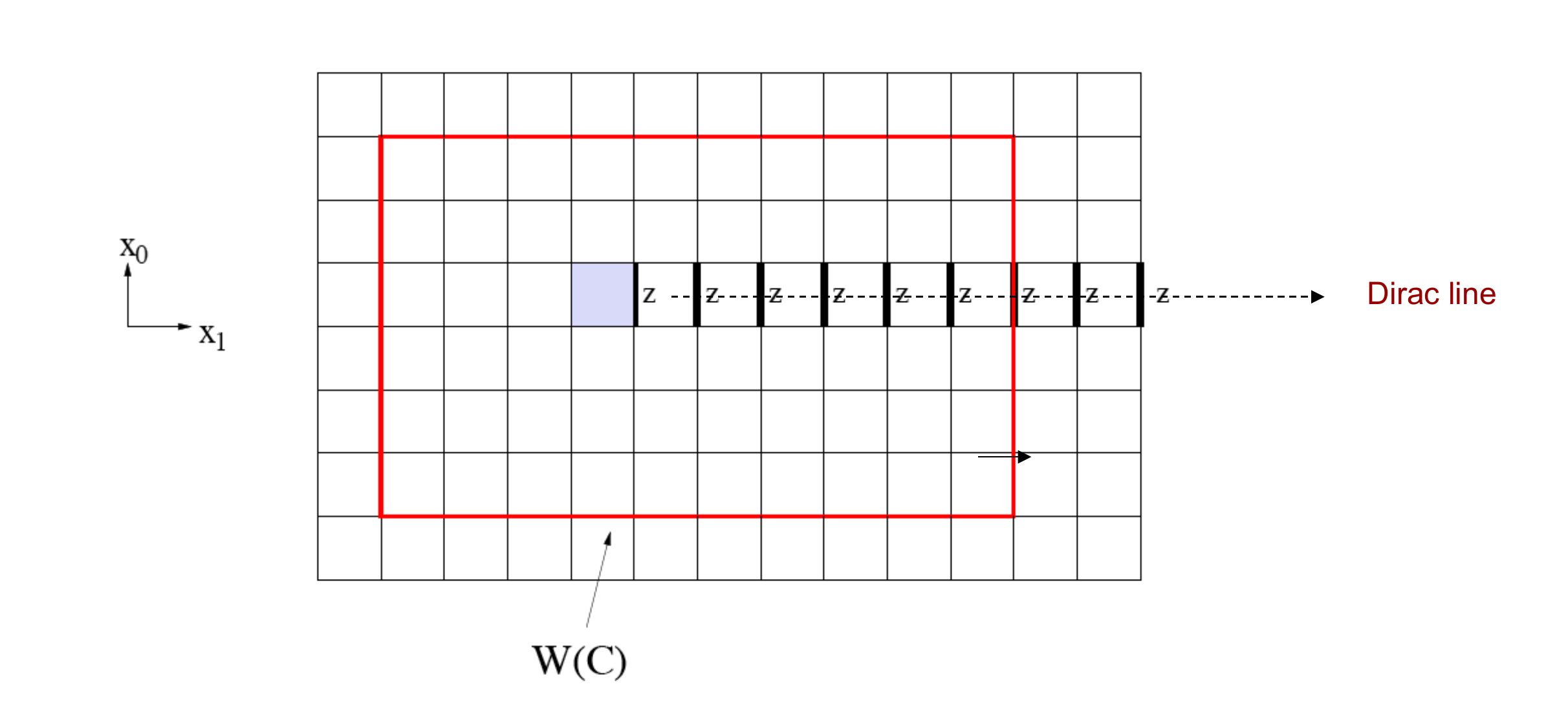}
}
\caption{(a) A global center transformation.  Each of the indicated links is multiplied by the same center element $z$; the lattice action is unchanged.  (b)  Creation of a thin center vortex.  The action is changed at the shaded plaquette, and a Wilson loop holonomy, linked to the vortex, is multiplied by the center element $z$.}
\label{squeeze}
\end{figure}

     Global center symmetry and center vortices are easiest to explain on the lattice.  The center of an SU(N) group consists of all the elements which commute with every element of the group, $z = \exp[i 2\pi (n-1)/N]\times \mathbbm{1} \in Z_N$.  Consider a transformation which multiplies every timelike link $U_0(\vx,t_0)$ on some timeslice $t=t_0$ by the same center element $z$, as shown in Fig.\ \ref{global}.  If the action is invariant under such a transformation, then the action is center symmetric. On the other hand, Polyakov lines on a periodic lattice are not invariant, but are multiplied by a center element.  This is why the VEV of a Polyakov line is an order parameter for center symmetry breaking.
     
     Creation of a center vortex can be thought of as an incomplete center transformation.  Suppose every timelike link 
$U_0(x_1\!\!>\!0,x_0)$ on a two-dimensional lattice is multiplied by the same center element.  Then the action is unchanged except at the shaded plaquette shown in Fig.\ \ref{cv3}.  This is a ``thin'' center vortex.  In two dimensions it is pointlike, in three dimensions (there is a line of shaded plaquettes coming out of the page) it is linelike, and in four dimensions a center vortex is surfacelike.  In any number of dimensions, a vortex can be topologically linked to a loop, and the effect of a vortex linked to a Wilson loop is to multiply the Wilson loop by a center element.  Of course, thin vortices are objects with a very high action density, and this action can be lowered by smearing out the vortex configuration.
These are called ``thick'' center vortices.  However, thin or thick, the effect of linking a center vortex to a large Wilson loop is the same.

\section{The Confinement Mechanism}
\label{mechanism}

    Let the gauge group be SU(2) for simplicity.  Consider a plane of area $L^2$ which is pierced, at random locations, by $N$ center vortices, and consider a Wilson loop of area $A$ lying in that plane.  Then the probability that $n$ of those $N$ vortices will lie inside the area $A$ is
\beq
          P_N(n) = {N \choose n}\left( A\over L^2 \right)^n \left(1 - {A \over L^2} \right)^{N-n}
\eeq
Each vortex piercing the Wilson loop contributes a factor of $-1$, so the vortex contribution to the Wilson loop is
\beq
     W(C) = \sum_{n=0}^N (-1)^n P_N(n) = \left(1- {2A \over L^2} \right)^N
\eeq
Now keeping the vortex density $\rho=N/L^2$ fixed, and taking the $N,L \ra \infty$ limit, we arrive at the Wilson loop area law falloff
\beq
W(C) = \lim_{N\ra \infty} \left(1 - {2\rho A \over N} \right)^N = e^{-2 \rho A}
\eeq
That is the vortex confinement mechanism in three lines \cite{Engelhardt:1998wu}.  I believe it is the simplest known.  The crucial assumption is that vortex piercings in the plane are random and uncorrelated, and this implies that vortices percolate throughout the spacetime volume.

\subsection{Semiclassical Considerations}

   Are there any theoretical reasons to believe that center vortex configurations might be present in the vacuum?  Center vortices,
unlike instantons (and their finite temperature cousins, the calorons) are definitely not saddlepoints of the classical action.  But matters are different at one loop, and the old work of Ambjorn and Olesen \cite{Ambjorn}, and later work by Diakonov and Maul 
\cite{Diakonov}, both suggest that center vortices are saddlepoints of the effective action at one-loop level.  There is also a simple argument based on lattice theory:   Consider a lattice action with traces in the fundamental (F) and adjoint (A) representations
\beq
         S =   \sum_{plaq} \bigg( \b_F  \tr_F[UUU^\dg U^\dg] + \b_A  \tr_A[UUU^\dg U^\dg] \bigg)
\label{Seff}
\eeq
Then if $\b_A \gg \b_F$ it is not hard to show that center vortices are saddlepoints of the action \rf{Seff} \cite{Greensite:2003bk}.   The effective lattice action at large distance scales surely contains the terms shown, as well as terms involving larger loops and higher representations.  But since zero N-ality loops fall off at large scales with a perimeter rather than an area law, the strong-coupling expansion tells us that the coefficients of zero N-ality loops (like $\b_A$) in the effective action will be be larger than the coefficients of non-zero N-ality loops (like $\b_F$).  In that case the same argument applies, and center vortices are saddlepoints of the effective action at large scales.

\section{Finding Center Vortices}
\label{find}

   The vortex mechanism is certainly the confinement mechanism in $Z_N$ gauge theories, since vortices are the only excitations.  Testing the idea in SU(N) gauge theories, via lattice Monte Carlo simulations, requires a method for locating thick vortices in thermalized configurations. The strategy is to map SU(N) configurations into $Z_N$ configurations, and use the thin vortices of the $Z_N$ lattice,
known at {\sl P-vortices}, to locate the thick vortices of the SU(N) lattice \cite{DelDebbio:1998luz}.  We first make a gauge transformation which brings the SU(N) link variables as close as possible to center elements, which is done by maximizing the quantity
\beq
            R = \sum_{x,\m} \mbox{Tr}[U_\m(x)] \mbox{Tr}[U^\dg_\m(x)]  \non
\eeq
This is known as ``maximal center gauge.''  We then project each SU(N) link to the closest $Z_N$ center element, e.g.\ for SU(2)
we set $ z_\m(x) = \mbox{sign Tr}[U_\m(x)]$, a procedure known as ``center projection.''  

\begin{figure}
\centering
\sidecaption
\includegraphics[width=0.5\textwidth]{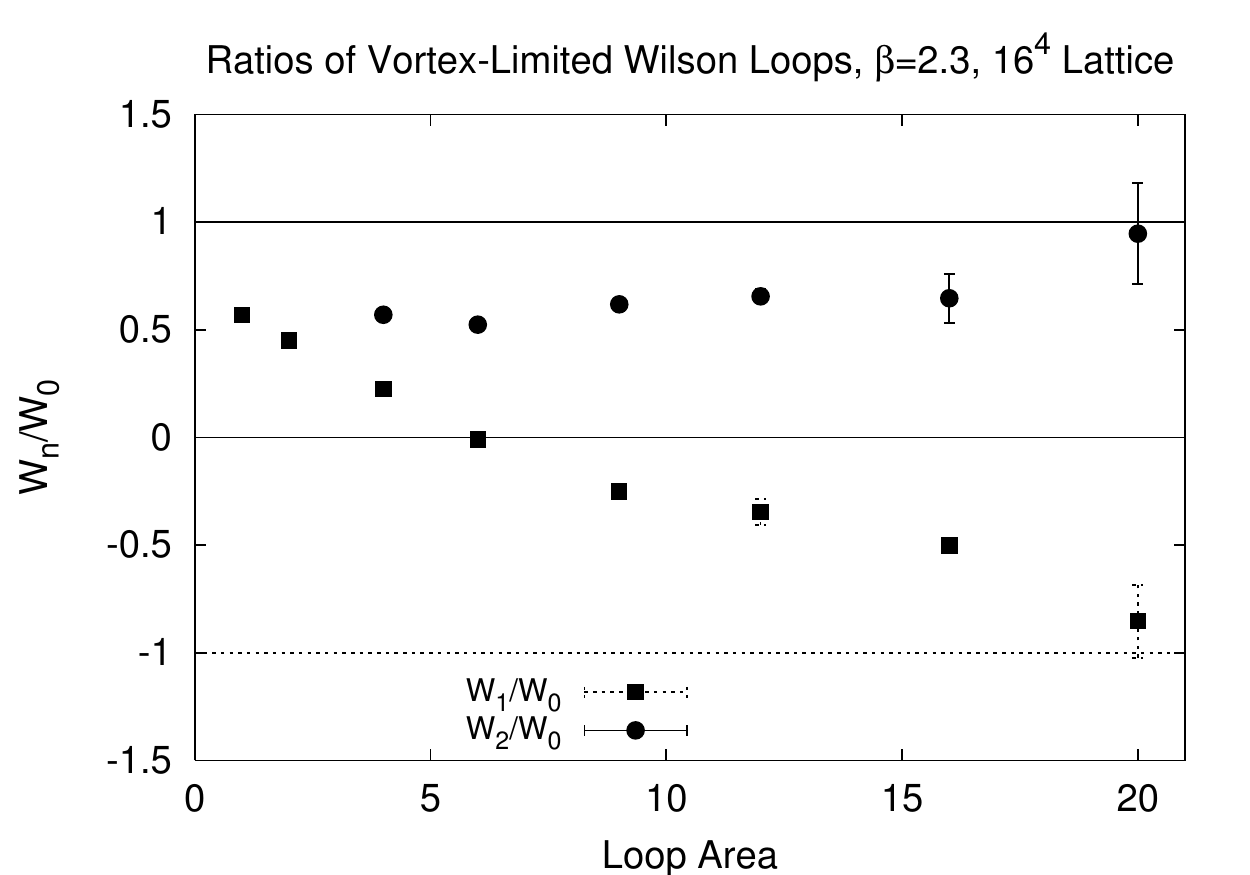}
\caption{Vortex-limited Wilson loops: a check that P-vortices locate center vortices (from \cite{DelDebbio:1998luz}).}
\label{Wn}
\end{figure}

   But do P-vortices on the projected lattice actually locate thick center vortices on the full lattice?  Define a ``vortex limited'' Wilson loop
$W_n(C)$  to be the VEV of Wilson loops on the SU(N) lattice with the condition that loop $C$ is linked to $n$ P-vortices on the projected $Z_N$ lattice.  Then if P-vortices locate thick center vortices of the unprojected lattice, we expect that in SU(2) gauge theory
\beq
           {W_n(C) \over W_0(C)} \ra (-1)^n  \non
\eeq
and this has been checked (see Fig.\ \ref{Wn}).  One also finds that the asymptotic string tension of vortex limited loops vanishes (since there are no fluctuations in sign due to variable linking numbers), and that SU(N) plaquettes whose location corresponds to P-vortices on the projected lattice have a substantially higher than average plaquette energy.  The correlation of P-vortices with gauge invariant observables is crucial to the statement that P-vortices do indeed locate thick center vortices on the unprojected lattice.

\section{Numerical Tests (Old)}

\begin{figure}[h!]
\centering
\subfigure[]  % caption for subfigure a
{   
 \label{temp1}
 \includegraphics[scale=0.3]{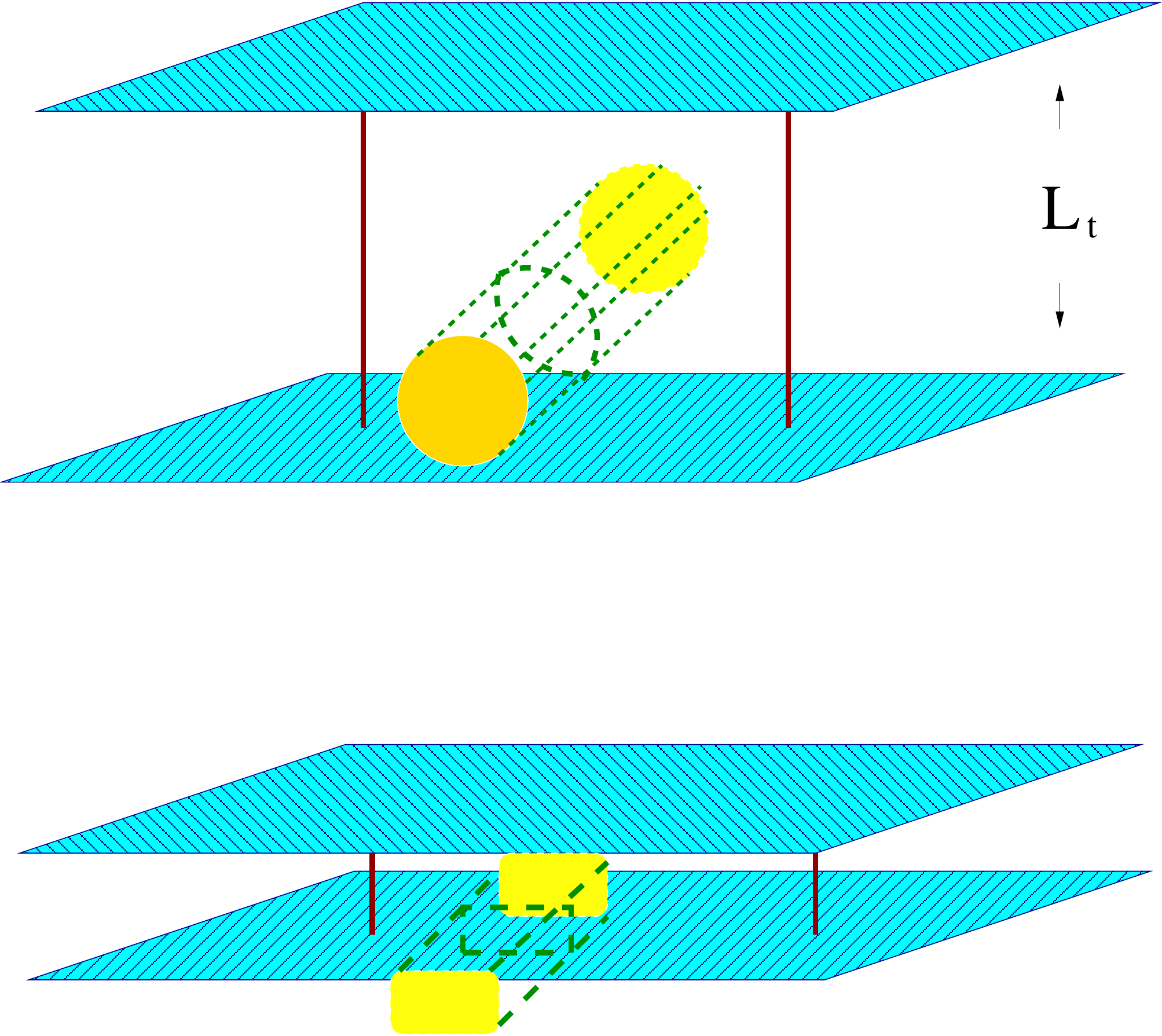}
}
\subfigure[]  % caption for subfigure a
{   
 \label{temp2}
 \includegraphics[scale=0.3]{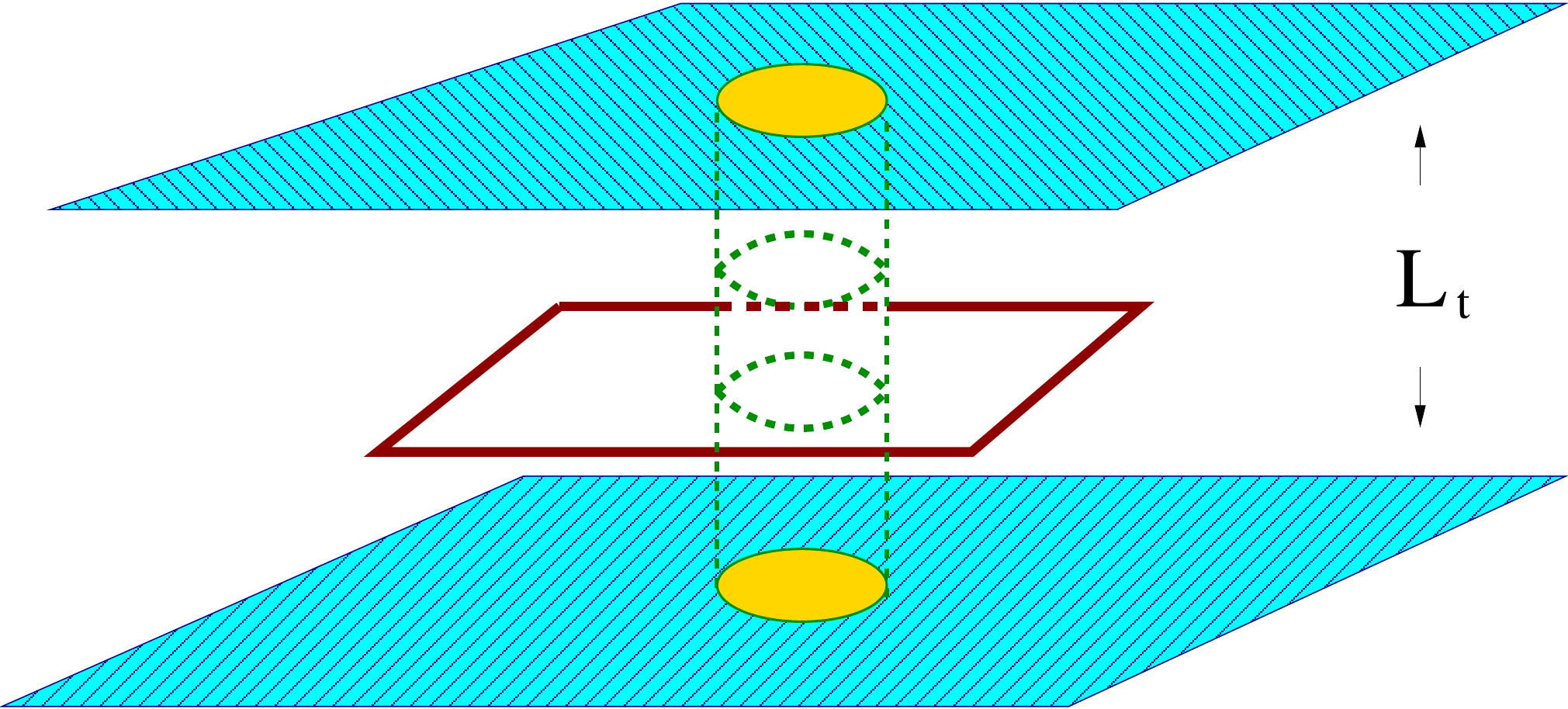}
}
\caption{(a) Vortices running in spacelike directions disorder Polyakov loops at low temperature (large $L_t$).  When vortices are
``squeezed'' at small $L_t$ their free energy rises, and they cease to percolate.  (b) Vortices running in the timelike direction disorder
spacelike Wilson loops.  The vortex thickness is not constrained by small $L_t$, and vortex free energy remains negligible.}
\label{squeeze}
\end{figure}

With a method in hand for finding thick center vortices, we can carry out various tests.  The following were carried out long ago, and I will quote only the results.  Details and figures can found in reviews \cite{Greensite:2011zz,Greensite:2003bk,Engelhardt:2004pf}, and in the cited references.
\begin{enumerate}

\item {\bf Center Dominance:}  How closely does string tension on the projected lattice approximate the asymptotic string tension
on the full lattice?  In SU(2) gauge theory, the agreement is quite good (90\% or better), although this is a little dependent on the details of the maximal center gauge fixing algorithm.  In SU(3) the agreement is not as good, with center projection resulting in about ${2\over 3}$ of the right answer.  Recently, however, the Adelaide group has found that the agreement in SU(3) improves considerably when the string tensions in the full and projected configurations are compared after some smoothing of each configuration
\cite{Trewartha:2015ida}.

\item {\bf Vortex Removal:}  Center vortices can be removed from a lattice configuration very simply, just by multiplying full links by the conjugate of projected links:
\beq
            U_\m(x) \ra U'_\m(x) = Z^*_\m(x) U_\m(x) \non
\eeq

\bigskip
This procedure leaves all local gauge-invariant observables untouched, except exactly at P-vortex locations, where
the action is actually increased.  The striking result, first noted by de Forcrand and D'Elia \cite{deForcrand:1999our}, is that when vortices are removed from the unprojected lattice, the asymptotic string tension drops to zero.

\item {\bf Scaling:} If P-vortices locate center vortices, and if these objects are physical rather than lattice or gauge-fixing artifacts, then their density should have a finite continuum limit. Vortex density deduced from the projected lattice is given by
\beq
          \rho = {\mbox{vortex area} \over \mbox{lattice volume}} {1\over a(\beta)^2} \non
\eeq
where $a(\beta)$ is the lattice spacing in physical units at lattice coupling $\beta$.  This quantity was shown to indeed have a finite continuum limit by Gubarev et al.\  \cite{Gubarev:2002ek}.

\item {\bf Finite Temperature:}  At high temperature the static quark potential goes flat, but spacelike Wilson loops still have an area law.
In the vortex picture these facts have a simple explanation, illustrated in Fig.\ \ref{squeeze}.  Center vortices have a finite thickness, which can be estimated from the vortex free energy as a function of lattice size \cite{Kovacs:2000sy,deForcrand:2001nd}.  Vortices running in the spacelike directions disorder Polyakov loops.  When the time extension $L_t$ is smaller than the diameter of the vortex (high temperature case), then spacelike vortices are ``squeezed" (Fig.\ \ref{temp1}),  their free energy increases, they cease to percolate, and the potential measured by Polyakov loop correlations goes flat.  In contrast, vortices running in the timelike direction disorder spacelike Wilson loops.  The vortex cross section is not constrained by a small extension in the time direction (Fig.\ \ref{temp2}), the vortex free energy of such vortices does not increase, and spacelike Wilson loops retain an area law even at high tempertures.  That, at least, is the picture.  It was verified in lattice simulations by the T\"ubingen group \cite{Engelhardt:1999fd}, who showed that vortex percolation in spacelike directions ceases above the deconfinement transition, but remains for vortices oriented in the time direction.

\item {\bf Topological Charge:}  P-vortex surfaces carry fractional topological charge at self-intersection points and so-called 
``writhing points''  \cite{Engelhardt:2000wc}.  Vortex surface models have been constructed which get observables such as 
topological susceptability about right \cite{Engelhardt:2010ft}.

\item {\bf Coulomb Confinement:}  There are strong connections between the Gribov-Zwanziger picture of confinement in Coulomb gauge, having to do with the density of eigenvalues of the Faddeev-Popov operator in infrared, and the vortex picture.  Upon vortex removal, the string tension of the color Coulomb potential vanishes, and the Faddeev-Popov eigenvalue density in the infrared resembles that of a free 
theory \cite{Greensite:2004ur}. 

\end{enumerate}

\section{Numerical Tests (New)}

    There are some recent results \cite{Trewartha:2015nna,Trewartha:2015ida} due to the Adelaide group of Kamleh, Leinweber, and Trewartha which I consider quite important, and will briefly review here (figures in this section are taken from \cite{Trewartha:2015nna}).  These authors begin by calculating the Landau gauge quark propagator using the overlap Dirac operator, for
\begin{itemize}
\item full (``untouched'') configurations
\item vortex-removed configurations
\item center projected after some smoothing steps (``vortex only'') 
\end{itemize}
and fit to the form
\bigskip
\begin{equation} 
S(p) = \frac{Z(p)}{ip\!\!\!/\ + M(p)} \non
\end{equation}
Smoothing is necessary for the center-projected configurations, because the overlap Dirac operator isn't suited to rough $Z_N$ configurations.  Smoothing is carried out in the full SU(3) manifold, not the $Z_N$ subgroup, so it is more appropriate to call the
result "vortex-only" rather than center-projected.

\begin{figure}[thb]
\centering
\subfigure[]{
\includegraphics[scale=0.3]{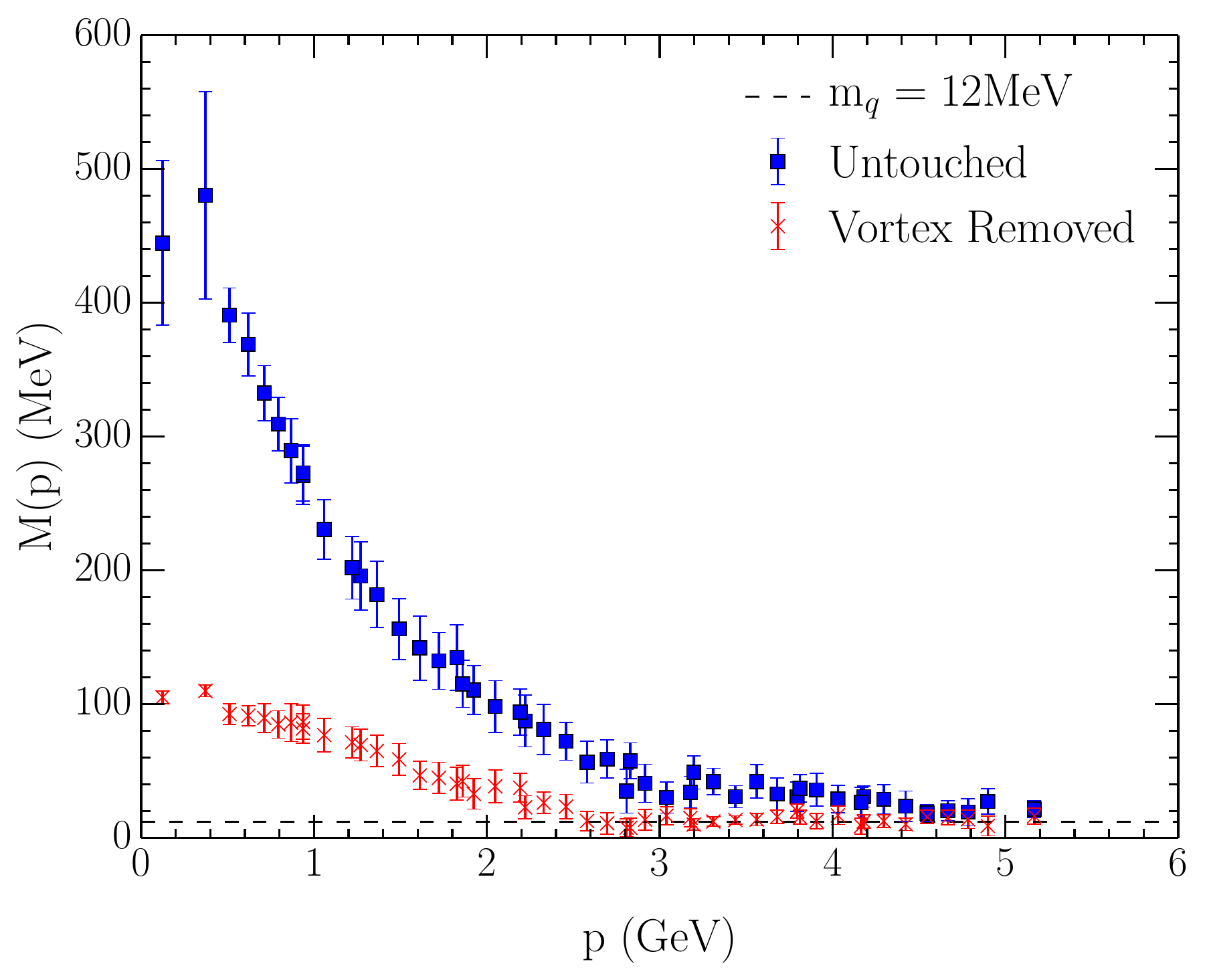}
}
\subfigure[]{
\includegraphics[scale=0.3]{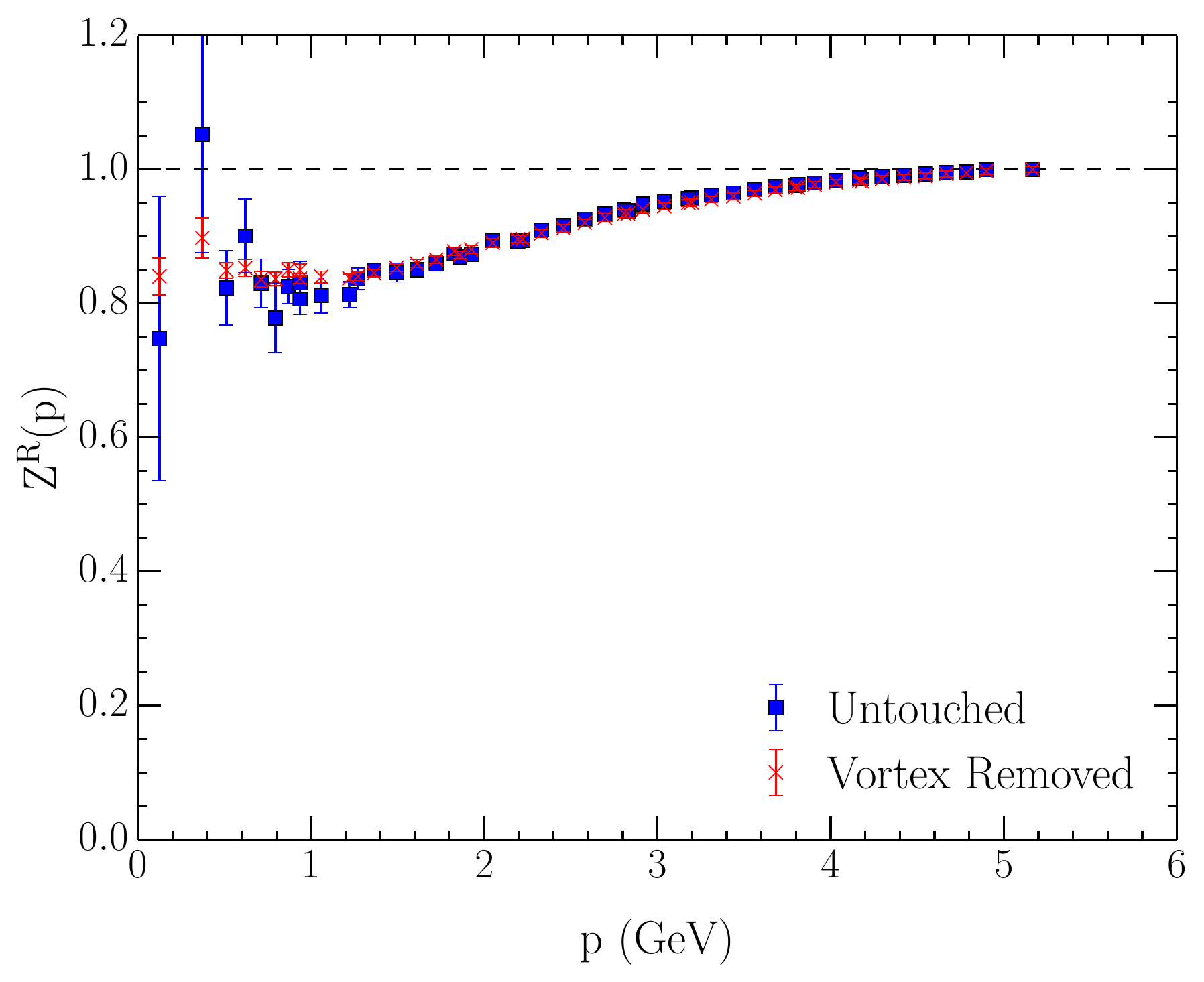}
}
\caption{The mass (a) and renormalisation (b) functions on the
  original (untouched) (squares) and vortex-removed (crosses)
  configurations.  Removal of the vortex structure from the gauge
  fields spoils dynamical mass generation and thus dynamical chiral
  symmetry breaking.}
\label{qprop1}
\end{figure}

\begin{figure}[thb]
\centering
\subfigure[]{
\includegraphics[scale=0.3]{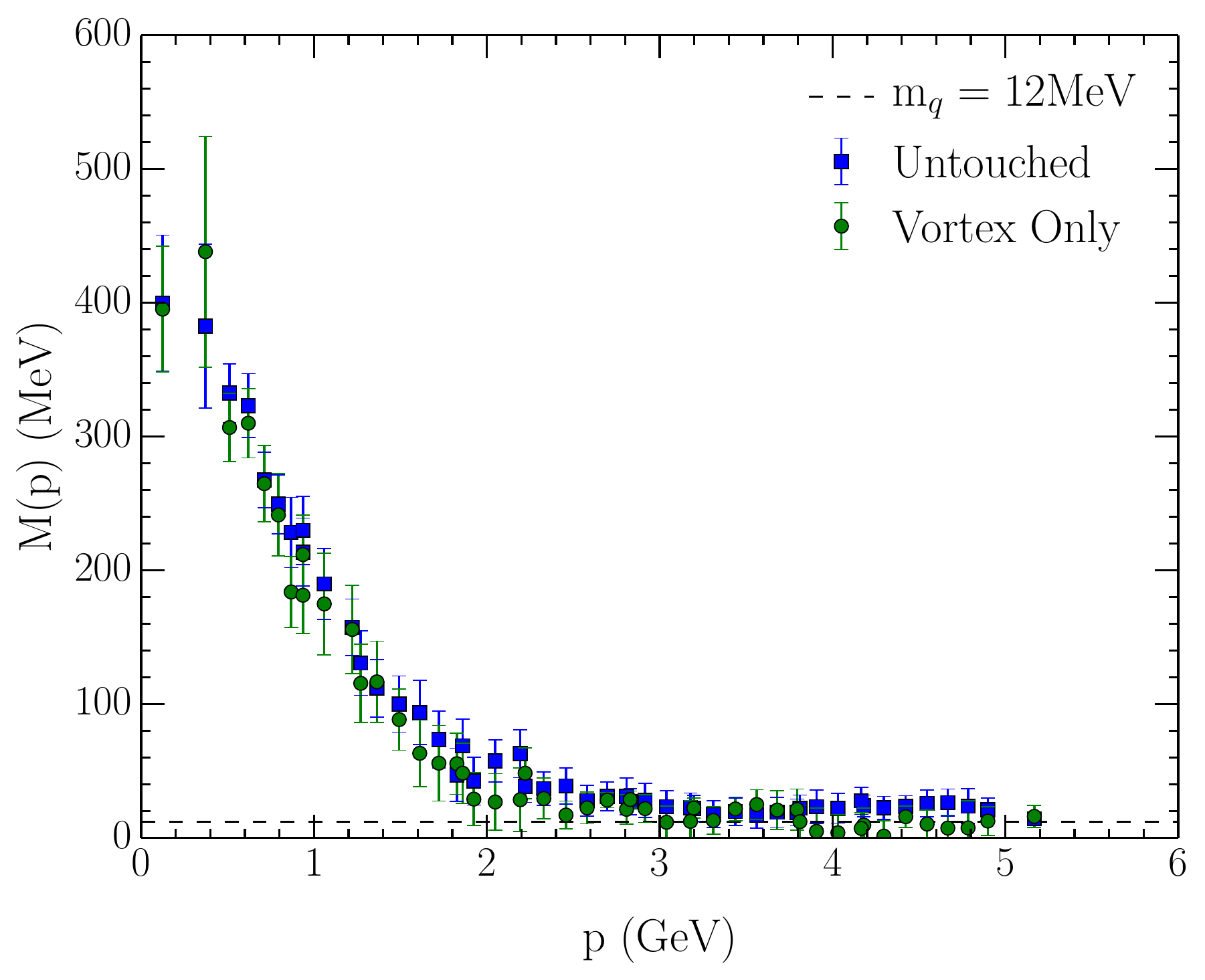}
}
\subfigure[]{
\includegraphics[scale=0.3]{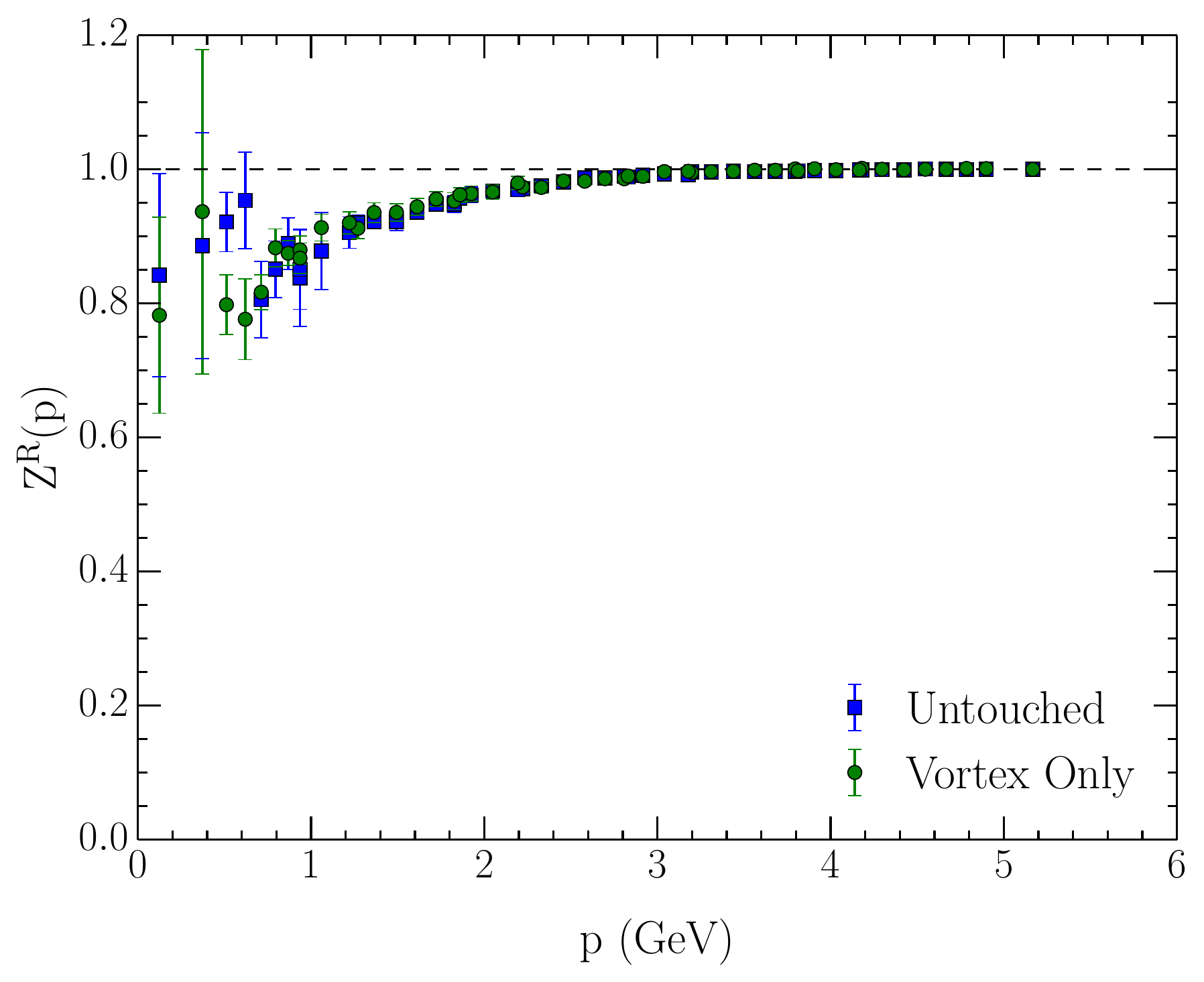}
}
\caption{The mass(a) and renormalisation (b) functions on the original (untouched) (squares)
  and vortex-only (circles) configurations after 10 sweeps of three-loop
  $\mathcal{O}(a^4)$-improved cooling, at an input bare quark mass of
  $12$ MeV.}
\label{qprop2}
\end{figure}

    A comparison of $M(p), Z(P)$ for the full and vortex removed configurations is shown in Fig.\ \ref{qprop1}.  Obviously vortex removal causes a severe reduction of the mass function, bringing it much closer to zero.  By contrast, $M(p),Z(p)$ for the full and vortex-only configurations, are virtually identical (Fig.\ \ref{qprop2}), and $M(p)$ is of course an effect due to chiral symmetry breaking. 

\begin{figure}[h!]
\centering
\sidecaption
  \includegraphics[width=0.4\textwidth]{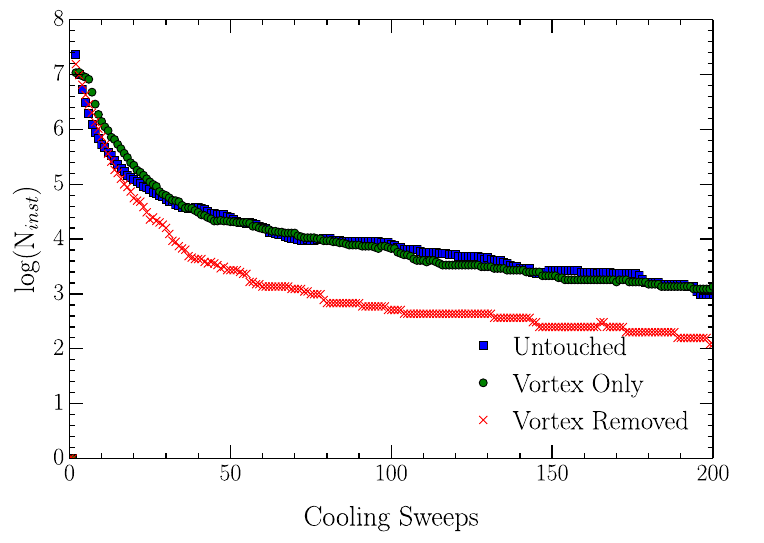}   
\caption{A logplot of the number of instanton-like objects per configuration found on untouched, vortex-only and vortex-removed ensembles as a function of O(a4)-improved cooling sweeps.}
\label{instanton}
\end{figure}

   The next question is topological charge.  After ten or so cooling sweeps it was found that maxima of the action in vortex-only configurations stabilized, and began to resemble classical instantons both in shape and corresponding topological charge density at the center.  A comparison of instanton number (on a logarithmic scale) vs.\ cooling sweeps in the full, vortex-only, and vortex-removed configurations is shown in Fig.\ \ref{instanton}.   Remarkably, the instanton content of the full and vortex-only theories is nearly identical, while the instanton density in vortex-removed configurations is drastically reduced by comparison.  Since instanton configurations appear to emerge from vortex-only configurations after a little cooling, the speculation is that vortices in some way contain the ``seeds'' of instantons.

   As for confinement, it was found that the string tensions of the full and vortex-only configurations were in close agreement after some cooling sweeps, while the asymptotic string tension vanished in the vortex-removed configurations.
 
   Finally,  the Adelaide group also computed low-lying hadron masses in vortex-only and vortex-removed ensembles, with the following
results \cite{Kamleh}:
\begin{itemize}

\item The vortex-only spectrum is qualitatively (and even quantitatively) very similar to full QCD.

\item The vortex-removed spectrum
   \begin{itemize}
      \item shows chiral symmetry restoration for light quarks; the
      pion is no longer a Goldstone boson;
      \item is a weakly-interacting theory of constituent quarks at heavy quark masses.
   \end{itemize}
\end{itemize}

    To summarize:  After a few cooling sweeps an ensemble of center vortices, extracted from thermalized lattice configurations, is essentially identical in all relevant infrared properties to an ensemble of full lattice configurations subjected to the same number of cooling sweeps, where the relevant properties are the string tension, chiral symmetry breaking, instanton content, and the spectrum.  I think it is hard to avoid the conclusion that it is the center vortex configurations which are driving these infrared effects both cases.   
   
\section{Double-Winding Loops and Abelian Confinement Mechanisms}

    Still, center vortices are not the only confinement mechanism on the market.  Monopoles and calorons/dyons are also popular.  In comparing and contrasting these ideas, I think that the example of multiple winding loops (or double-winding loops in SU(2) gauge theory) are useful observables to consider \cite{Greensite:2014gra}. 
    
\begin{figure}[h!]
\centering
\subfigure[]  % caption for subfigure a
{   
 \label{ks504}
 \includegraphics[scale=0.3]{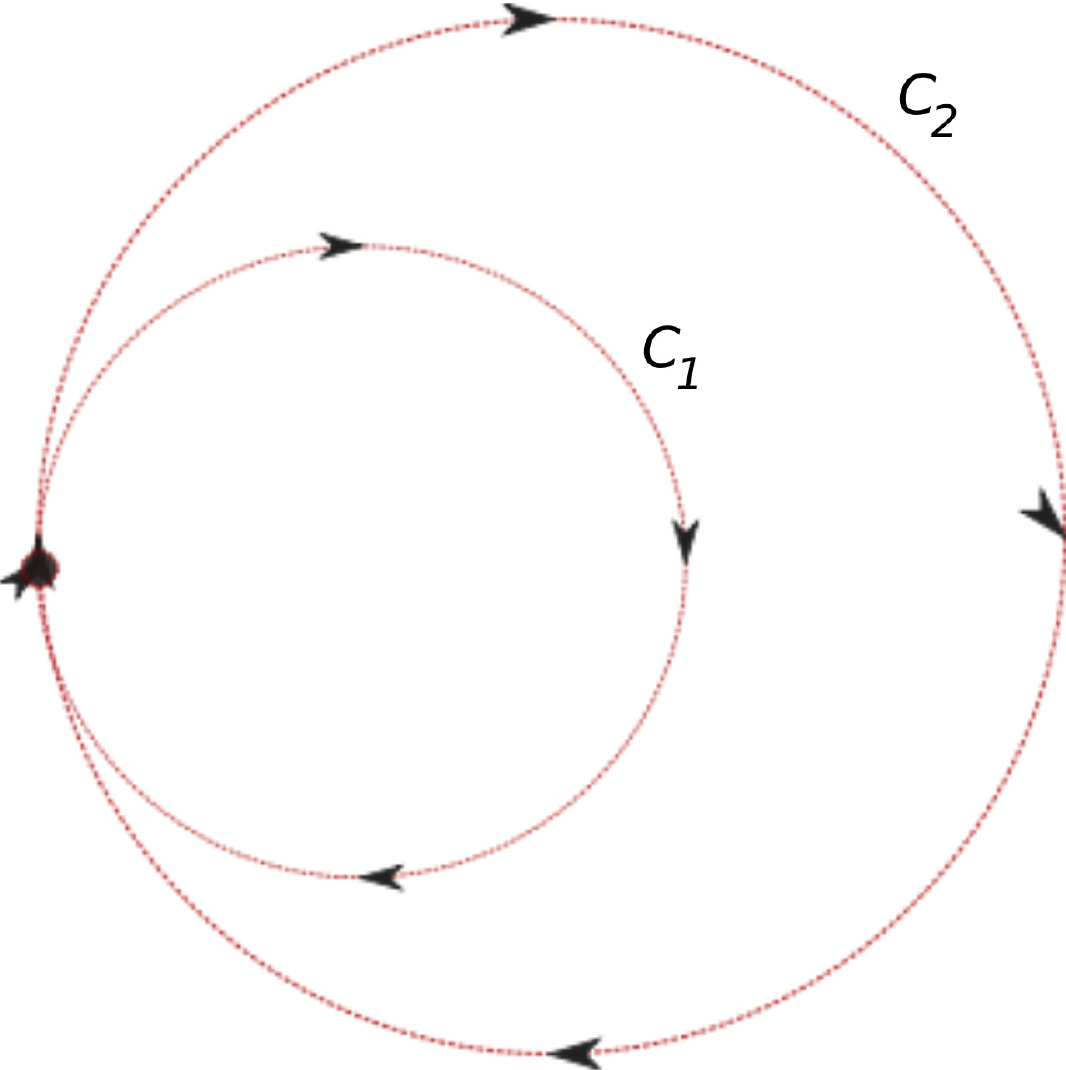}
}
\subfigure[]  % caption for subfigure a
{   
 \label{kk504}
 \includegraphics[scale=0.3]{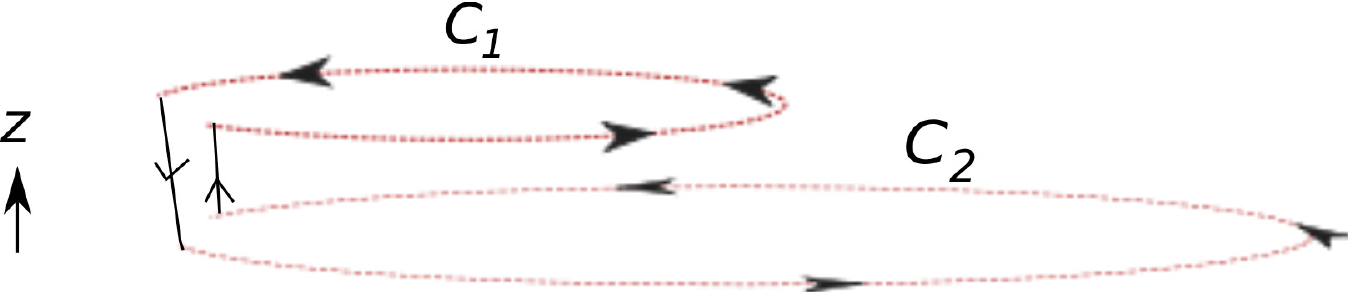}
}
\caption{(a) A double winding loop, which runs once around contour $C_1$, and once around the
coplanar loop $C_2$. (b) A ``shifted'' double winding loop, in which contours $C_1$ and $C_2$ lie in planes parallel to
 the $x-y$ plane, but are displaced from one another in the transverse direction by distance $\delta z$, and are connected by lines running parallel to the $z$-axis.}
\label{dwind}
\end{figure}

    A double-winding Wilson loop is a closed loop which winds once around the closed contour $C_1$ and once (winding in the same direction) around the closed contour $C_2$.  Contours $C_{1,2}$ are either coplanar or, for technical reasons, in parallel planes but slightly displaced from one another in an orthogonal direction (see Fig.\ \ref{dwind}).  The corresponding minimal areas are $A_1$ and $A_2$ respectively.  What are the predictions of abelian confinement mechanisms, i.e.\  dual superconductivity and the monopole/dyon plasma, as compared to the center vortex theory?
Abelian confinement mechanisms emphasize the role of gluons in the Cartan subalgebra of the gauge group; the other gluons, sometimes referred to as ``W-bosons,'' are thought to be very massive, irrelevant to the infrared physics, and are typically ignored for this reason.  We will also ignore them for the time being.

     In a dual superconductor model, by the dual Meissner effect, there will be two electric flux sheets bounded by $C_1$ and $C_2$.
If we think of the loops as lying parallel to the $z-t$ plane,  then in any time slice there are two flux tubes parallel to the $z$-axis.
This leads, in SU(2) gauge theory, to a sum-of-areas falloff of the double-winding loop
\beq
          W(C) \sim \exp[-\s (A_1+A_2)] 
\eeq
In the case of a monopole or dyon plasma, following Polyakov \cite{Polyakov:1976fu} and Diakonov and Petrov \cite{Diakonov:2007nv}, there is a soliton spanning $A_1$, and
another soliton spanning $A_2$ (for simplicity we take the orthogonal separation between $A_{1,2}$ to be greater than the width of the 
soliton).  This leads again to a sum-of-areas law.  Intuitively, the reason is that $C_{1,2}$ behave like two current loops inserted into a monopole plasma.  The magnetic field generated by the currents are screened by two monopole-antimonopole sheets along the areas
$A_1$ and $A_2$.

    In contrast, the prediction of the center vortex mechanism is a difference-of-areas law
\beq
          W(C) \sim \exp[-\s |A_1-A_2|] 
\eeq
This is because, in SU(2) gauge theory, a vortex linked to both loops has no effect; i.e.\ two factors of $-1$ are $+1$.  Fluctuations in sign due to vortices can only come about due to vortices linked to the larger of the two loops but not the smaller, and this is what leads to the difference-of-areas law.  The numerical evidence is very clearly in favor of the difference-of-areas law \cite{Greensite:2014gra}.

\subsection{What about the W's?}
 
    In order to obtain the difference-in-areas law, abelian confinement mechanisms must appeal to screening by $W$-particles,
which in SU(2) are doubly charged under the abelian subgroup.  But this raises the following question:  what happens if we imagine integrating out the $W$'s?  Only the abelian field is left.  What are then the confining configurations, which are able to give the 
proper difference-of-area falloff for the double-winding loops?  The vacuum can't simply be a dual superconductor or monopole/dyon plasma, because these give the wrong answer.

   I would suggest that after integrating out the $W$'s, the result will be a center vortex vacuum.  There is, in fact, one example where this can be shown to occur.  
   
   Consider a fixed modulus $|\rho|=1$ double-charged matter field in U(1) lattice gauge theory
\bea
  Z &=&  \int D\rho  D\theta_\m ~ \exp\left[ \beta \sum_p \cos(\theta(p)) 
  + \oh \lambda \sum_{x,\m} \left\{\rho^*(x)
              e^{2i\theta_\m(x)} \rho(x+\widehat{\m}) + \mbox{c.c}
          \right\} \right] 
\eea
with $\beta \ll 1$ (confinement), $\lambda \gg1$.
In the absence of matter fields we have confinement via a Coulomb gas of magnetic
monopoles, and string tension is proportional to charge.  But with those matter fields, even charged Wilson loops $\langle U(C)^{2n} \rangle$ have zero string tension, and odd-charged loops have the same string tension.  
{\sl This changes the monopole plasma to an ensemble of $Z_2$ center vortices.} To see this, go to unitary gauge $\rho=1$, and make the field decomposition
\beq
       \exp[i\theta_\m(x)] = z_\m(x) \exp[i\tth_\m(x)]   ~~~
\mbox{where} ~~~
        z_\m(x) \equiv \mbox{sign}[\cos(\theta_\m(x))] 
\eeq
and
\bea
     Z &=&   \prod_{x,\m} \sum_{z_\m(x)=\pm 1}
           \int_{-\pi/2}^{\pi/2} {d\tth_\m(x) \over 2\pi}
 \exp\left[ \beta \sum_p Z(p) \cos(\tth(p)) +
                     \lambda \sum_{x,\m} \cos(2\tth_\m(x)) \right]
\eea
One can easily show, for ${\beta \ll 1,~\lambda \gg 1}$, that
\beq
      \Bigl\langle \exp[in\theta(C)] \Bigr\rangle \approx \langle Z^n(C) \rangle
                        \Bigl\langle \exp[in\tth(C)] \Bigr\rangle \non
\eeq
with
\bea
     \langle Z^n(C) \rangle &=& \left\{ \begin{array}{cl}
         \exp[-\s A(C)] & n \mbox{~odd} \cr
               1        & n \mbox{~even} \end{array} \right. 
\non \\
      \Bigl\langle \exp[in\tth(C)] \Bigr\rangle
             &=& \exp[-\m n^2 P(C)] \non
\eea
where $Z(C)$ is the product of $z_\m(x)$ link variables around the loop $C$.
This establishes that the confining fluctuations are thin $Z_2$ center vortices, identified by the
$z_\m$ variables in unitary gauge.

\begin{figure}[h!]
\centering
\sidecaption
\includegraphics[width=7cm,clip]{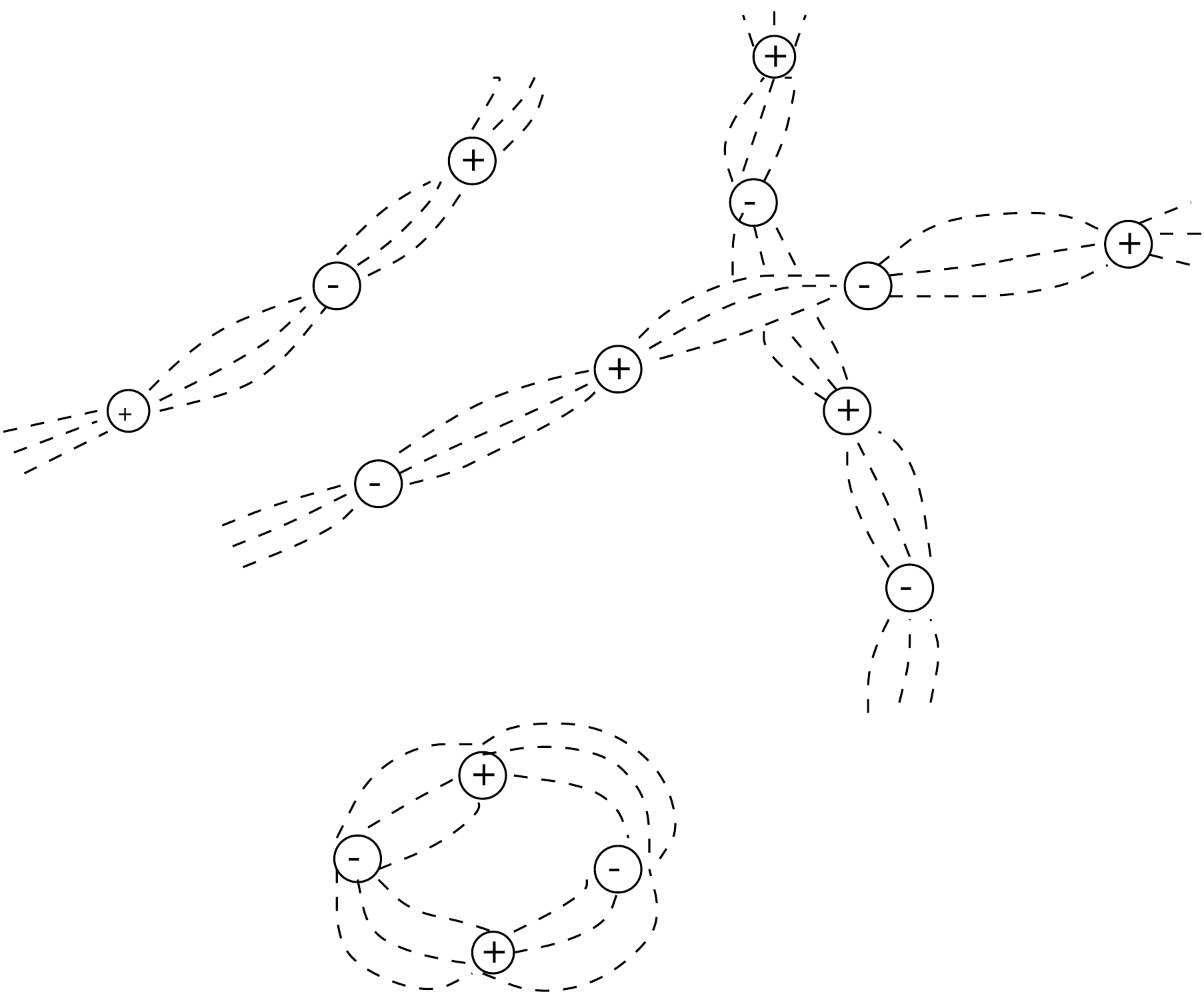}
\caption{Hypothetical collimation of monopole-antimonopole flux into center vortex flux tubes.}
\label{chain}       % Give a unique label
\end{figure}

   One way to picture what has happened is that the monopoles and antimonopoles have lined up into monopole-antimonopoles
chains, with the corresponding fields collimated into a $Z_2$ vortex, something like Fig.\ \ref{chain}.   So if we assume that monopoles or dyons are part of the story, then this is probably the result obtained after integration over the $W$-fields.  In fact, this kind of collimation has been seen in computer simulations of SU(2) gauge theory, on abelian-projected lattices \cite{Ambjorn:1999ym}.

\section{Conclusions}

   Center vortices provide a plausible and well-motivated mechanism for confinement, deconfinement, chiral symmetry breaking, and the generation of topological charge.  It is more than a model.  Center vortices are found in thermalized lattices obtained from lattice Monte Carlo simulations, and it is found that 
\begin{itemize}
\item vortex density scales according to asymptotic freedom,
\item vortex-only configurations account, more-or-less accurately, for the observed asymptotic string tension, quark propagator, 
instanton density, and the general qualitative features of the hadron spectrum, while
\item vortex-removed configurations have none of these properties.
\end{itemize}

   That's the good news.  The bad news is that vortex configurations, as thick surfaces in four Euclidean dimensions, do not lend themselves to an analytic treatment.   So while I believe the vortex theory is correct, I also think it will probably not help us to solve QCD.
The method which really does solve QCD is the lattice Monte Carlo method, and yet using this method to compute, e.g., the hadron spectrum, does not necessarily supply much physical insight.  It may be that, like it or not, solving QCD and understanding QCD are simply different problems.

% BibTeX or Biber users please use (the style is already called in the class, ensure that the "woc.bst" style is in your local directory)
% \bibliography{name or your bibliography database}

\bibliography{vortex}
%Journal Author, Journal \textbf{Volume}, page numbers (year)
%Book Author, \textit{Book title} (Publisher, place, year) page numbers
% etc
%\end{thebibliography}

\end{document}